\documentclass[fleqn,10pt]{wlscirep}
\usepackage[utf8]{inputenc}
\usepackage[T1]{fontenc}
\usepackage[capitalise]{cleveref}
\usepackage{subcaption}
\usepackage{physics}
\usepackage{booktabs, multirow}
\title{QuKAN: A Quantum Circuit Born Machine approach to Quantum Kolmogorov Arnold Networks}

\author[1,3,*]{Yannick Werner}
\author[1,2]{Akash Malemath}
\author[3]{Mengxi Liu}
\author[3]{Vitor Fortes Rey}
\author[1,3]{Nikolaos Palaiodimopoulos}
\author[1,3]{Paul Lukowicz}
\author[1,2,3]{Maximilian Kiefer-Emmanouilidis}
\affil[1]{Department of Computer Science and Research Initiative
QC-AI, RPTU Kaiserslautern-Landau, Kaiserslautern,
Germany}
\affil[2]{Department of Physics, RPTU Kaiserslautern-Landau,
Kaiserslautern, Germany}
\affil[3]{Embedded Intelligence, German Research Center for Artificial Intelligence (DFKI),
Kaiserslautern, Germany}

\affil[*]{mun60zor@rptu.de}


\begin{abstract}

Kolmogorov Arnold Networks (KANs), built upon the Kolmogorov Arnold representation theorem (KAR), have demonstrated promising capabilities in expressing complex functions with fewer neurons. This is achieved by implementing learnable parameters on the edges instead of on the nodes, unlike traditional networks such as Multi-Layer Perceptrons (MLPs). However, KANs potential in quantum machine learning has not yet been well explored.  
In this work, we present an implementation of these KAN architectures in both hybrid and fully quantum forms using a Quantum Circuit Born Machine (QCBM). We adapt the KAN transfer using pre-trained residual functions, thereby exploiting the representational power of parametrized quantum circuits. In the
hybrid model we combine classical KAN components with quantum subroutines, while the fully quantum version the entire architecture of the residual function is translated to a quantum model. We demonstrate the feasibility, interpretability and performance of the proposed Quantum
KAN (QuKAN) architecture. 
\end{abstract}

\begin{document}

\flushbottom
\maketitle

\thispagestyle{empty}

\section*{Introduction}

As proposed by Liu et al\cite{liu2024kan}, the generalization of the original Kolmogorov Arnold representation theorem \cite{schmidt2021kolmogorov} to arbitrary width and depth can yield a model capable of holding its ground to classical Multi Layer Perceptron \cite{haykin1994neural,hornik1989multilayer} (MLP) in terms of interpretability and accuracy. The structure of the generalization of the original theorem into the KAN relies on trainable residual function that are represented as linear combinations of a set of basis functions. This strong resemblance between the linear combinations and the superposition representation of quantum mechanical wavefunctions motivates the implementation of the KAN as a quantum model. The potential benefits of a quantum approach to KAN go beyond structural analogy. Quantum systems inherently support the parallel evaluation of multiple functions via superposition, enabling operations on exponentially large Hilbert spaces\cite{nielsen2010quantum,schuld2018supervised}. Although, previous work exist on quantum KANs, it is largely preliminary.   Thus, QKAN\cite{ivashkov2024qkan} is limited to translating the residual function as a unitary representation and has not yet been evaluated on its training performance. An initial concept for a Variational Quantum KAN (VQKAN) has been presented in \cite{wakaura2024variational, wakaura2025adaptive, wakaura2025enhanced}. However, despite being more robust against noise, the enhanced version \cite{wakaura2025enhanced}, where the residual functions are evaluated through the construction of a tiled matrix using sum operators, introduces an exponential overhead when the number of layers is increased.

In this work, we propose a feasible and simple quantum generalization of KAN via Quantum Circuit Born Machines \cite{liu2018differentiable} (QCBM). These models learn to generate target probability functions using the probability interpretation of quantum physics given by the Quantum Born rule \cite{stoica2025born}. In this paper we go beyond probabilistic sampling (which is how QCBMs are usually used) and propose encoding entire residual functions into quantum states via weighted superpositions. Using projective measurements we can evaluate multiple functions at once allowing us to represent classical KAN residual functions as a trainable quantum circuit by including a division of the computational basis into labelling and position. We then propose an even more general form of the network that includes superposition interpretation of all the parts of the residual functions, yielding an effective fully quantum residual. \\
Finally, we demonstrate the models capabilities and performance on simple datasets with Binary Classification and function approximation and compare to sparsely available results from an Enhanced VQKAN \cite{wakaura2025enhanced}, and to a trainable version of QKAN \cite{ivashkov2024qkan}. Furthermore, we also present results from comparable  Variational Quantum Circuits, a rigid grid pyKAN and two- and four layered MLP where we evaluate the \textit{make\_moons}  and Iris\cite{unwin2021iris} dataset.

\section*{Methods}
In this section, we present the implementation of the Hybrid Quantum KAN. Our model leverages the structural resemblance between the definition of B-Splines\cite{de1978practical} as linear combinations of functions and the general representation of an arbitrary quantum state as a linear combination of basis states. We provide a brief overview of the QCBM, the superposition-based function approximation, the readout mechanism, and the construction of both single and combined QuKAN residual functions.
 
\subsection*{Quantum Circuit Born Machines}
Quantum Circuit Born Machines are a promising tool in the regime of unsupervised generative learning of quantum circuits due to their high expressive power \cite{du2020expressive}. The training utilizes the probabilistic interpretation of the wave function of a quantum state in a given representation, as described by the Quantum Born rule \cite{brumer2006born,stoica2025born}. It states that the probability of measuring a quantum system in a particular state is equal to the squared magnitude of the amplitude resulting from the projection of the wavefunction onto that state. This stands in contrast to the idea of Boltzmann machines that leverage thermal distributions \cite{hinton2007boltzmann}. Given a target function as a dataset of independent samples, the QCBM uses projective measurements in the computational qubit basis. The outcome probability distribution is then expected to resemble the discretized target data. For our purposes, the quantum circuit always starts in the state $|0\rangle^{\bigotimes n}$, representing the $n-$qubit computational basis state $|00\dots 00\rangle$. It consists of strongly entangling layers\cite{schuld2020circuit}, composed of parametrized rotation gates and nearest neighbour controlled NOT gates. The loss is computed from the measured output, and the circuit parameters are updated using gradient based learning \cite{liu2018differentiable}. 
\begin{figure}[!h]
    \centering
    \includegraphics[width=0.95\linewidth]{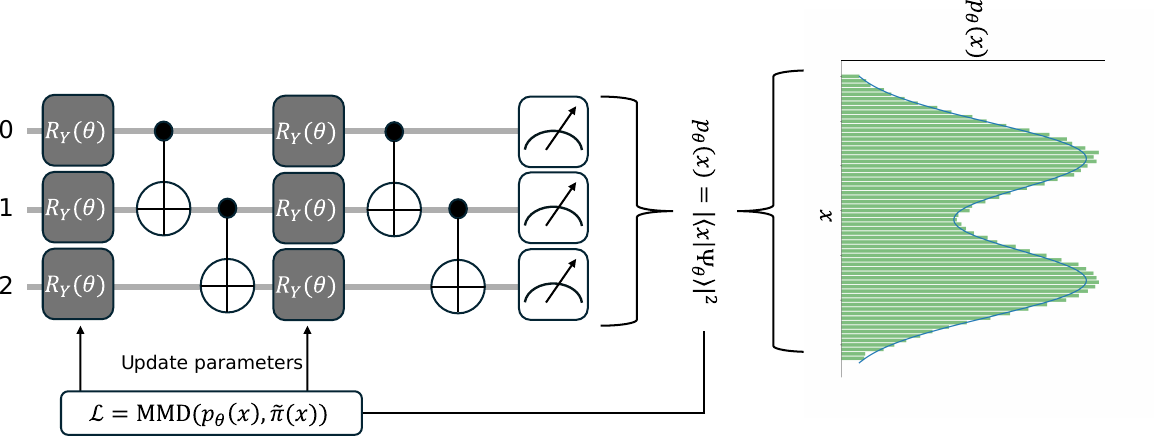}
    \caption{Sketch of the Quantum Circuit Born Machine learning algorithm: Starting from the state $|0\rangle^{\bigotimes n}$ we process through a quantum circuit containing strongly entangling layers, so parametrized rotations as well as CNOT gates. The full computational basis is measured via projective measurements and the squared maximum mean discrepancy (MMD) loss is calculated. Note that the comparison here goes along the normalized target distribution $\tilde{\pi}(x)$, since the total probability across the computational basis is constrained to sum to one. After training we can perform another run and generate the target. }
    \label{fig1}
\end{figure}
In order to work in low dimensional feature spaces we map the data using Kernel methods as proposed by Gretton et al \cite{gretton2006kernel}. As the QCBM uses projective measurements we want to highlight that the input features are given only at the final stage, defined by the position indicated by the binary string obtained from the readout. This resembles the probability amplitude of the corresponding computational basis state, see Fig.~\ref{fig1}. 
In the following we will use the QCBM as a form of amplitude embedding for the corresponding spline basis functions. One can immediately use Mottonen encoding \cite{1mottonen2005}, which introduces a large overhead due to its exponential circuit depth scaling with the amount of qubits.

\subsection*{Superposition Distribution Learning}
\begin{figure}[!h]
    \centering
    \includegraphics[width=0.9\linewidth]{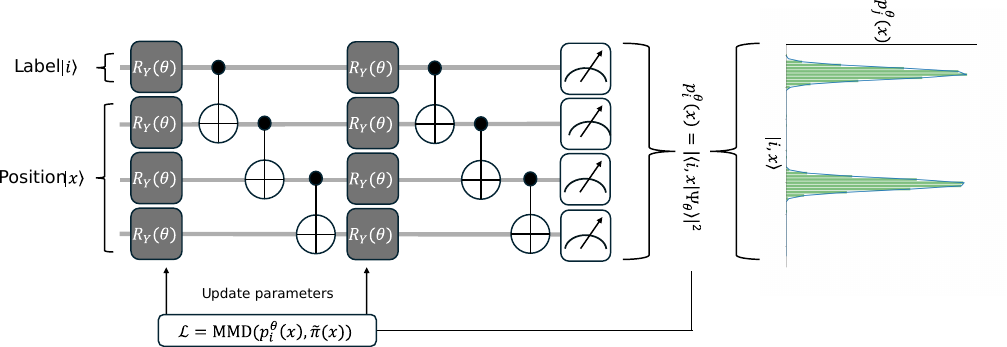}
    \caption{Sketch of the training of the QCBM for parallel superposition learning of two target functions. The process follows the same algorithm as shown in figure [\ref{fig1}]. The difference lies in the comparison for the optimization. Here we compare to the state representing both functions as given in the Kronecker basis denoted by $\tilde{\pi}(x)$. }
    \label{fig:QCBMSuperpos}
\end{figure}

For later purposes we want to generalize the learning capabilities of a QCBM using the quantum superposition principle. This allows a quantum system to represent multiple classical target functions simultaneously in contrast to conventional models. Specifically, we show that a single quantum state can encode a collection of discretized probability distributions, each corresponding to a different target, within its structure. To this aim we introduce a division of the computational basis of the circuit into labelling qubits (denoted by $i$) and position qubits (denoted by $x$). In quantum mechanical terms, the overall state of a system composed of these two subsystems is expressed as a tensor product (denoted by $\bigotimes$) of their individual states. This can be written as
\begin{equation}
    |\Psi\rangle = \sum_{i=1}^{N_i}c_i|i,\psi_i\rangle = \sum_{i=1}^{N_i}c_{i}\sum_{k=0}^{N_k} d_{ik}\ket{\psi_k},
    \label{eq:register}
\end{equation}
where $|i,\psi_i\rangle = |i\rangle \bigotimes|\psi_i\rangle$ is the tensor or Kronecker product \cite{sakurai2020modern} of the labelling and position subspaces. Here, $c_i \in \mathbb{C}$ are the amplitudes which are initialized with $1/\sqrt{N_i}$. The index $i$ runs over the number of trained target (labeled) states $N_i$ that exist in the Kronecker basis and $d_{ik} \in \mathbb{C}$ runs over $N_k$ states of the corresponding position register of the target state. 
To extract information about a particular target probability distribution $p_j(x)$, which we later interpret as the target function $f_j(\cdot)$ at a given input feature point $x$, we perform a projective measurement\cite{nielsen2010quantum} in the computational basis over the combined label and position qubits.  
The outcome probability of measuring a particular basis state is given by
\begin{equation}
    p^\theta_j(x) = |\langle j,x| \Psi^{\theta}\rangle|^2 =  |\langle j,x| \sum_i c_{i}|i,\psi_i^\theta\rangle|^2 = |\sum_i c_{i} \psi^\theta_i(x) \delta_{ij} |^2 = |c_{j}\psi^{\theta}_j(x)|^2= |c_{j}\sum_k d_{jk}(\theta)\psi_k(x)|^2,
    \label{eq:pretrain}
\end{equation}
where $\theta$'s are computed in the pre-training phase.
Note that taking the absolute squared of the complex valued amplitudes maps them to real valued probabilities. The $\delta_{i,j}$ denotes the Kronecker delta that represents the orthonormality of the computational basis on which we encode. The effect of this projection is to isolate the squared amplitude which, guaranteed by the Born rule\cite{brumer2006born,stoica2025born}, gives the probability of measuring a specific label and position pair, see Fig.~\ref{fig:QCBMSuperpos}. Crucially, this approach allows the QCBM to be trained just as the original formulation by adjusting the the weights of the parametrized unitaries such that the measurement matches the empirical data. The key difference lies in the encoding that utilizes quantum benefits, namely superposition. 

\subsection*{Hybrid KAN Residual Functions}
In this section we are going to introduce a hybrid formulation of the Quantum KAN residual function utilizing the superposition distribution learning described in the previous section. As this directly maps onto the classical KAN architecture, we will first provide a summary of the classical counterpart. 

\subsubsection*{Summary of classical KAN}

The Kolmogorov--Arnold representation theorem is a foundational result in multivariate function theory, stating that any continuous function of multiple variables can be represented as a superposition of continuous univariate functions and addition. 
Formally, the theorem states that for any continuous multivariate function on a bounded domain $f: [0,1]^n \rightarrow \mathbb{R}$, there exist continuous functions $\phi_i$ with a single variable and $g_j$ such that:

\begin{equation}
    f(x_1, \dots, x_n) = \sum_{j=1}^{2n+1} g_j\left(\sum_{i=1}^{n} \phi_i(x_i)\right)
    \label{eq:kan_thre}
\end{equation}

This result, originally proved by Kolmogorov and later refined by Arnold \cite{kolmogorov1957representations,arnold2009functions}, has profound implications in the field of approximation theory. 
It guarantees that multivariate continuous functions can be constructed using only univariate function compositions and additions, without requiring explicit multivariate non-linearities.

KANs\cite{liu2024kan} build upon Kolmogorov--Arnold representation theorem.
While there are only two-layer non-linearities and a small number of terms $(2n+1)$ in the hidden layer according to this theorem, the authors generalized the network to arbitrary widths and depths by defining a single KAN layer $\phi(x)$ in the following way:
since the function $\phi_i(x_i)$ in \cref{eq:kan_thre} is a univariate function, it can be parametrized as a B-spline Curve $spline(x)$, with learnable coefficients $c_i$ of local B-spline basis functions $B_i(x)$ as shown in \cref{eq:bspline}, where $c_i$ is the trainable parameters. 

\begin{equation}
    spline(x) = \sum_{i}c_iB_i(x)
    \label{eq:bspline}
\end{equation}

Theoretically, $spline(x)$ can be implemented using the KAN layer $\phi_i(x_i)$. A residual architecture was designed to enhance its optimization. Consequently, the KAN layer was defined as shown in \cref{eq:kan_layer}, where $w_b$ and $W_s$ are trainable weights, retained in the original implementation to control the overall magnitude.

\begin{equation}
    \phi(x) = w_b \text{SiLU}(x) + w_s \sum_i c_i B_i(x), \quad \mathrm{with} \quad
 \text{SiLU}(x) = \frac{x}{1+\text{e}^{-x}},
 \label{eq:kan_layer}
\end{equation}
where SiLU is the Sigmoid-weighted Linear Unit.
In traditional neural networks such as MLPs, each layer computes affine transformations followed by fixed element-wise non-linearities (e.g., ReLU or tanh). 
While these architectures are known to be universal approximators under certain conditions, they often require large numbers of neurons or layers to approximate complex functions effectively.
However, KAN replaces the fixed scalar weights between neurons with learnable univariate functions (B-splines). Instead of each edge carrying a scalar weight, it carries a learnable function $\phi(x)$, enabling the network to directly approximate the decomposition described in \cref{eq:kan_thre}. This allows KANs to express more complex functions with fewer neurons and deeper theoretical grounding.
We want to emphasize here that the definition of the B-spline part in the residual function already has great resemblance to the superposition structure of a quantum mechanical wavefunction. The goal of the next section is to see now how we can combine the QCBM representation of a wavefunction in position space to the learning scheme of a classical KAN residual function.

\subsubsection*{Quantum representation of the residual functions}
With the proposition of superposition based distribution learning via QCBM methods, we have demonstrated the possibility of encoding multiple classical functions into the probability distribution generated by the measurement statistics of a state. Building upon the previously introduced  label-position register decomposition in the computational basis, we now apply this framework to train the network on a predefined set of discretized B-spline basis functions given by the Cox-de-Boor recursion\cite{de1978practical}. These functions denoted by $B_i(x)$ form the building block of the classical KAN residual function, which takes the form
\begin{equation}
    f(x) = \sum_{i=1}^N \tilde{c}_i B_i(x),
\end{equation}
where $\tilde{c}_i\in \mathbb{R}$ are the classical scaling coefficients of the linear combinations and $B_i(x)$ is the $i$-th spline evaluated at input $x$. In the quantum formulation we encode the evaluation of the basis functions into the (normalized) amplitudes of a quantum state. 
\begin{equation}
    \ket{f}=\sum_{i=1}^{N_i} c_{i}|i\rangle|\beta_i\rangle,\quad \mathrm{and} \quad f_i(x)=\bra{i,x}\ket{f}, \quad \beta_i(x)=\bra{x}\ket{\beta_i}
    \label{eq:pf(x)}
\end{equation}
where $i$ indicates the label, $x$ denotes the position in the corresponding qubit register and $c_{i}$ represents the trainable amplitudes of the $\beta_i$ states which are pre-trained by the QCBM, see Eq.~(\ref{eq:pretrain}). The encoding corresponds to a superposition over basis functions at fixed positions weighted by trainable amplitudes $c_{i}$ that can be optimized via the application of parametrized unitary operations to the state. 
\begin{align}
   p_f(x)&= \sum _j |\bra{j,x} \ket{f}|^2= \sum _j | \bra{j} \sum_i c_i \beta_i(x)\ket{i}|^2 = \sum _i | c_i \beta_i(x)|^2 
\end{align}
As already indicated here, the training optimization of the amplitude parameters will only take place on the labelling part of the register adjusting how much each basis function is weighted. To retrieve the function value $f(x)$ we measure the position register in state $|x\rangle$ over all label projections. Note that this is equal to the sum of the weighted evaluations over all basis functions. To preserve the correct magnitude of the function approximation, we recognize that a normalization factor introduces downscaling of the amplitudes due to the probabilistic nature of quantum states \cite{sakurai2020modern}. However, since the normalization scaling is a constant factor, we can correct it either during post-processing or by absorbing it into the training objective. In particular, the classical coefficients can be reconstructed as 
\begin{equation}
    f(x)\sim p_f(x) , \mathrm{when}\quad  \tilde{c_i} = |c_i|^2,\quad \mathrm{and} \quad B_i(x) = |\beta_i(x)|^2
\end{equation}
ensuring that the learned distribution matches the target functions amplitude structure after rescaling.  
\begin{figure}[!h]
    \centering
    \includegraphics[width=0.7\linewidth]{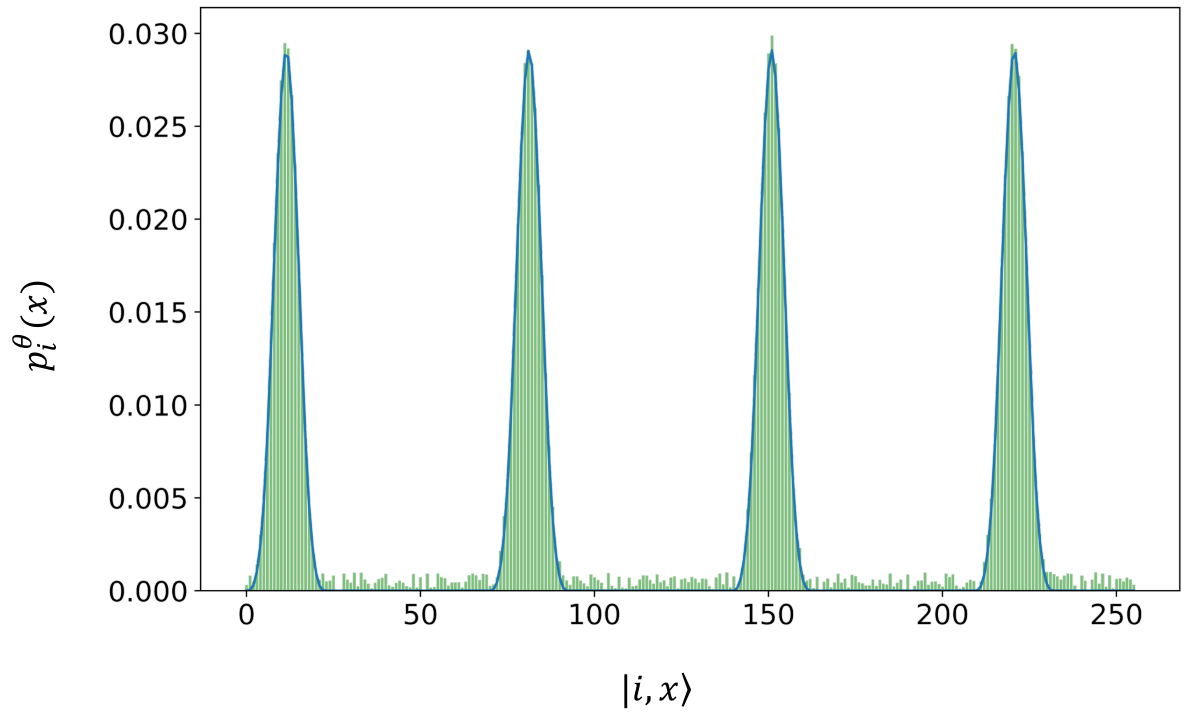}
    \caption{QCBM probability output for normalized B-spline basis functions on discretized input interval}
    \label{fig3}
\end{figure}

\subsubsection*{Hybrid QuKAN residual function}
To understand how information propagates through the QuKAN architecture, we begin by analysing the processing of a single hybrid residual function. This unit combines a Quantum Function Evaluator (QFE), trained to approximate a set of pre-trained basis functions, with a classical non-linear transformation. This is analogous to the architecture of the classical KAN.

The first step is to discretize the input data. This step is necessary due to the finite resolution of quantum registers, which restricts the number of distinguishable input values encoded into the position qubits by the exponentially large Hilbert space. Let $n_x$ denote the number of available qubits of the position register. Then the input feature space is partitioned into $2^{n_x}$ equally distanced points between the minimum and the maximum of the input range. Formally, if $X$ is the set of input features, then
\begin{equation}
    X = \{x_0,x_1,...,x_{2^{n_x}-1}\}, \,\,x_i = \min(X) + i\Delta x, \,\,\,\,\Delta x = \frac{\max(X)-\min(X)}{2^{n_x}-1}.   
\end{equation}
For any given input feature we determine the nearest discretized point by 
\begin{equation}
    x_{\text{meas}} = \text{argmin}_{x\in X}|x-x_{\text{input}}|,
\end{equation}
and assign it to that position in our position register which determines where we are going to perform the projective measurement for the readout.

We initialize our quantum circuit with a QCBM pre-trained quantum state representing a set of evaluations of B-spline basis functions as described in the previous section. The initial state is given as an equal superposition over the labelling register. During forward propagation the labelling qubits that index and weight the basis functions are passed through multiple parametrized entangling layers, analogous to classical weight training KANs. These trainable gates are set to optimize the coefficients in the linear combination of the basis functions, tailored to the given task. The position qubit register remains fixed and only comes into play by determining the position of the projective measurement. This measurement projects the total state onto the pre-determined $x_{\text{meas}}$ for all labels $j$ separately as presented in Eq.~(\ref{eq:pf(x)}). Optionally, now the obtained probability can be upscaled again.

\begin{figure}[!h]
    \centering
    \includegraphics[width=0.9\linewidth]{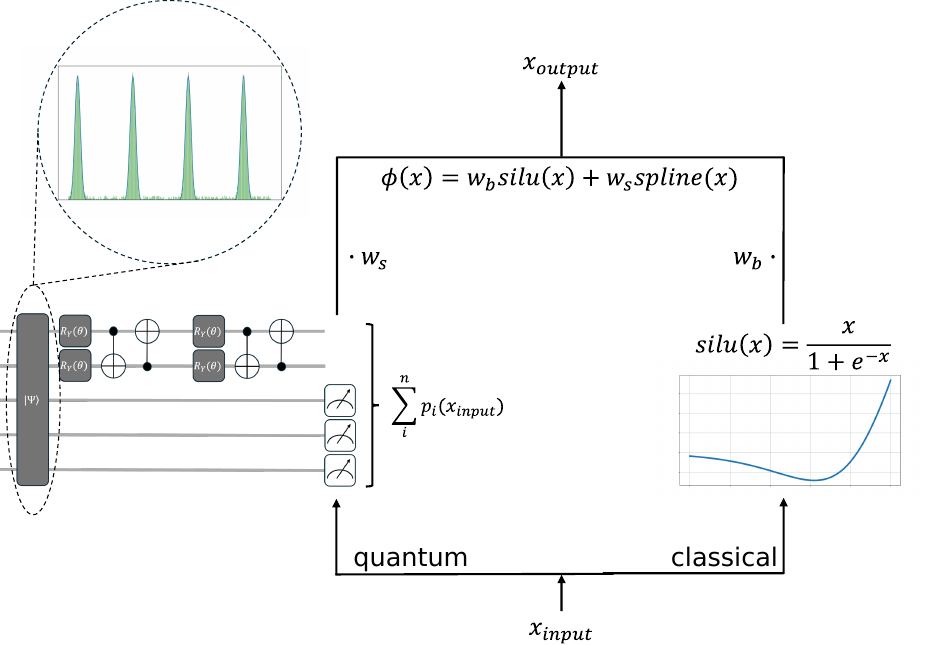}
    \caption{Architecture of a single Hybrid QuKAN residual function. The input data gets processed through a classical (the SiLU function) and a quantum transfer. The quantum transfer is based in the pre-trained QCBM Spline function encoding and optimizes the prefactors of the linear combination. Both get weighted and summed up to produce the output. }
    \label{fig4}
\end{figure}

In parallel to that, the unpreprocessed input is processed through a classical non-linear activation, such as (in analogy to the classical KAN) the SiLU. The final output of the residual function is obtained by summing over the quantum and classical part with the introduction of trainable scaling parameters as 
\begin{equation}
    f_{\text{Residual}}= w_f p_f(x) + w_s \text{SiLU}(x).
\end{equation}
A schematic of this process can be seen in Fig.~\ref{fig4}. 
With the definition of the structure of a single QuKAN residual function, we can extend this construction to form a full network. The architecture mirrors the structural paradigm of the KAR as introduced in the generalized KAN\cite{liu2024kan}, combining residuals into feed-forward networks of arbitrary width and depth.

\subsection*{Full Quantum KAN}
Using the QCBM for the pre-training it is possible to encode an arbitrary amount of functions into a superposition state. There are essentially two limitations that arise from the nature of the Born rule, i.e. that the readout of the circuit is a probability. It has to be normalized and positive. However it is possible to keep the normalization factor as well as a shift from a (reasonable) function as numerical constants, to be applied after the encoding. To see that the original definition of a residual function is essentially not different from the superposition we used for the Splines in terms of basis functions. The SiLU is also weighted and added which allows us to encode it into the superposition of the Splines as another added factor. The fully quantum residual function that consists of a weighted superposition of spline basis functions as well as the SiLU. This changes the form of the hybrid residual function such that the classical part is absorbed into the quantum side. We evaluate in parallel not only the splines at a given input feature using projective measurement but also include the SiLU via pre-training a QCBM on the complete superposition of the classical KANs residual function. 

\subsection*{Summary of the methods}
In this section, we have introduced the architecture and core mechanism of the QuKAN as a hybrid model that integrates quantum-enhanced function representation. By leveraging the superposition principle we have demonstrated how a single quantum state trained via a QCBM can encode multiple discretized B-Spline basis functions simultaneously. This followed a decomposition of the computational register into labelling and position qubits allowing us to evaluate multiple functions with projective measurements of the position register. This hybrid architecture maintains the functional interpretability and compositional power of classical KANs, while introducing quantum native parallelism and probabilistic expressibility.

\section*{Results}
In this section we will present a few benchmarks of the performance of the hybrid quantum KAN in both classification and function approximation tasks. Regarding classification tasks we evaluated our method in two datasets, namely the two moons dataset as provided in the scikit-learn\cite{scikit-learn}  \textit{make\_moons} function  and the Iris\cite{unwin2021iris} dataset. For function approximation, we tested our approach with two variable functions, including some evaluated in Wakaura et al\cite{wakaura2025enhanced}.

\subsubsection*{Binary Classification}

For the classification tasks we will compare our model's performance to that of a Variational Quantum Classifier (VQC) using different encoding strategies. Since the Quantum KAN architecture is close to a VQC with pre-training and ancillas, we also show that the pre-training, namely the encoding of the spline basis functions, has a positive effect. 

\begin{figure}[!h]
    \centering
    \includegraphics[width=1.1\linewidth]{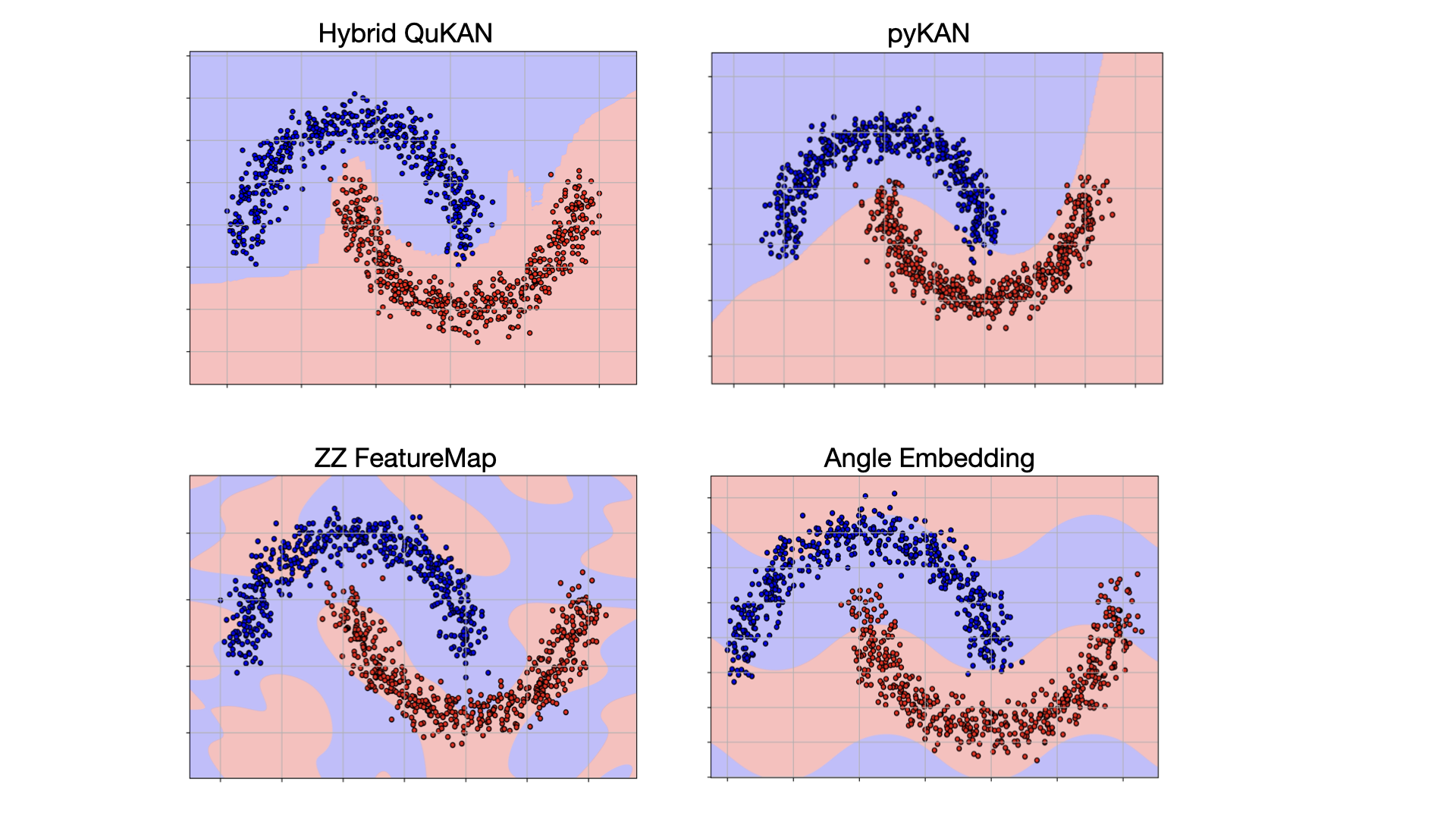}
    \caption{Decision boundaries on the moons dataset (with noise=0.1) for different models}
    \label{fig2}
\end{figure}

In order to compare the proposed Quantum KAN architecture to other quantum methods we chose binary classification as a first benchmark. We set up the network consisting of 2 layers and initialize each residual function as an equal superposition of 4 splines of degree 2. For the moons dataset we use a total of 1000 examples for training and the same amount of samples for testing. Both sets are independently sampled using a noise level of $0.1$. To check the stability of the model as well as compare it to different quantum classifier models, we present in \cref{fig2} the decision boundary after training our proposed QuKAN, together with that of other models. 
Namely, we compare the performance and accuracy of our hybrid QuKANs with those of competing quantum models, as well as the classical pyKAN. For pyKAN we introduced a rigid grid and limited the maximum number of splines to 4 in order to ensure a fair comparison with the QuKAN architecture. For quantum methods, we mainly focus on Variational Quantum Classifiers\cite{Schuld_2020,farhi2018classificationquantumneuralnetworks} that perform classification by processing the encoded data through a strongly entangling layer architecture \cite{Schuld_2020}, followed by the evaluation of the expectation value of an observable on a single qubit. We also include different embedding strategies to enhance the performance of the quantum models, namely amplitude\cite{1mottonen2005} and angle embedding\cite{khan2024beyond}. For amplitude embedding we also include ancilla qubits to enable a meaningful comparison, as the QuKAN implementation is essentially a Parametrized Quantum Circuit (PQC) that incorporates both pre-training and ancilla qubits. For the angle embedding\cite{lloyd2020quantum} we include a ZZ-feature map \cite{lloyd2020quantum,suzuki2020analysis}. Finally, for the QKAN model proposed by Ivashkov et al \cite{ivashkov2024qkan, QKAN_Implementation_vanH} we implemented a simple autograd based optimization algorithm of their proposed transfer function to be able to compare to our proposed QuKAN architecture. The classification performance can be seen in \cref{tab1} while in \cref{fig2} we present the predicted decision boundaries for the moons dataset for the 4 different models. It is easy to see that QuKAN and pyKAN predict similar boundaries, however the one predicted by pyKAN is smoother. Since the performance of full quantum models is highly sensitive to their embedding strategies \cite{lloyd2020quantum,hwang2024quantum,helstrom1969quantum}, this becomes a limiting factor in their effectiveness in the case of Angle Embedding, including ZZ feature maps, as well as Amplitude Embedding. Furthermore, we note that VQC approaches with trainable observables have demonstrated improved performance on the moons dataset \cite{chen2025learning}, however, we leave comparison to those methods as future work and focus on simpler model architectures that are closer in complexity to our proposed method.

\begin{table}[!ht]
    \centering
    \begin{tabular}{|l|l|l|}
    \hline
        \textbf{Model} & \textbf{Mean Test Accuracy Moons} & \textbf{Mean Test Accuracy Iris}\\ \hline
        QuKAN & $\mathbf{97.94\%\pm0.14\%}$ & $\mathbf{100\%\pm0\%}$ \\  \hline 
        Rigid grid pyKAN & $97.74\%\pm0.34\%$ & $97.49\%\pm 0.014 \% $\\ \hline
        MLP (2 layers) & $86.52\%\pm0.93\%$&$81.14\%\pm 0.33\%$ \\ \hline
        MLP (4 layers) & $\mathbf{99.74\%\pm0.15\%}$ & $\mathbf{100\%\pm0\%}$\\ \hline
        VQC (Amplitude Embedding) &$84.78\%\pm0.001\%$  & $60.00\% \pm 0\%$\\ \hline
        VQC (Amplitude Embedding + Ancillas) & $83.96\%\pm0.001\%$ & $60.00\% \pm 0\%$\\ \hline
        VQC (Angle Embedding) & $80.18\%\pm0.004\%$ & $63.00\% \pm 0.04\%$\\ \hline
        VQC (ZZ FeatureMap) & $81.21\%\pm0.007\%$ & $66.99\% \pm 0.05\%$\\ \hline
        QKAN & $84.06\%\pm 0.005\%$ & $\mathbf{100\%\pm0\%}$\\ \hline
    \end{tabular}
    \caption{Comparison of the Mean Test Accuracy for different Machine Learning Models for the Make Moons (with noise=$0.1$) and Iris dataset. All models are initialized with an comparable amount of parameters and trained for 20 epochs. For better comparison of QuKAN and pyKAN we scaled pyKAN down to a rigid grid and only two layers as well as 4 splines per residual function.
    For QKAN we used 2 hidden layers of width 3 alongside the input and output layers with Chebyshev polynomials\cite{mason2002chebyshev} up to degree of 3 for each. All models are evaluated over 4 different seeds and the mean test accuracy is presented. The Accuracies for different noise levels in the dataset can be found in the Supplementary Material.}
    \label{tab1}
\end{table}
As presented in table [\ref{tab1}] the QuKAN and the rigid grid pyKAN show similar performance for both test sets. In the case of the moons dataset our QuKAN shows higher mean accuracies over different parameter initializations than the QKAN by Ivashkov et al.\cite{ivashkov2024qkan}, while their performance is similar for Iris. The QuKAN outperforms all the VQC methods with different embeddings for the tested datasets.

\subsubsection*{Function Regression}

\begin{figure}[!h]
    \centering
    \includegraphics[width=0.9\linewidth]{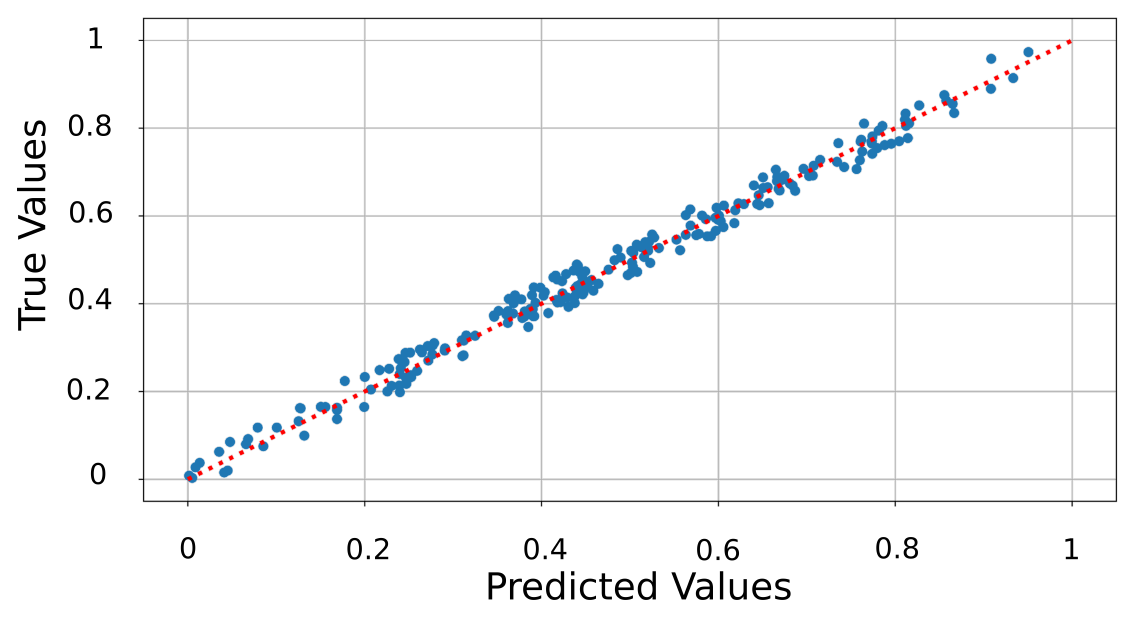}
    \caption{Comparison of the true against the predicted values for the regression task of the QuKAN. The red line indicates perfect prediction. }
    \label{LinFuncFig}
\end{figure}

As demonstrated in the original generalization of the KAN by Liu et al \cite{liu2024kan}, KANs are particularly well-suited for function regression tasks due to their structured and interpretable composition of basis functions. In this section we want to show how the Quantum KAN can be used to fit a multivariate function. We choose 
\begin{align}
    f(x_1,x_2) = 2x_1 -3x_2 +1
\end{align}
defined over the input domain $x_1,x_2\in[0,1]$. This function serves as a controlled benchmark to test the approximation capability of the model with limited number of parameters. We initialize the QuKAN model with two residual layers and limit it to 4 Splines basis function per residual. The result is visualized in Fig.~\ref{LinFuncFig}. 

To further evaluate the regression capabilities of our QuKAN model, we compare it to results reported in an implementation of a Quantum Kolmogorov Arnold Network proposed by Wakaura et al\cite{wakaura2025enhanced}. In this paper the sum of absolute distances between predicted and true function values is reported. We compare the models on the regression task of the function
\begin{equation}
    f(x_0,x_1) = \ln\left(\frac{x_0}{x_1}\right)
\end{equation}
and optimize via Mean Squared Error loss. We start by calculating the Average, median, minimum and maximum of the sum of absolute distances of the predicted value to the true values for a batch size of 250 as shown in table [\ref{table1}]. For comparison to the model proposed by Wakaura et al, we set the train set to 10 samples and the test set to 50 samples. The results are summarized in table [\ref{table4}]. In this case we also include training on the proposed full QuKAN architecture. 
\begin{table}[!ht]
    \centering
    \begin{tabular}{|l|l|l|l|l|}
    \hline
        \textbf{Model} & \textbf{Sum Abs. Dist. Avg.} & \textbf{Sum Abs. Dist. Med.} & \textbf{Sum Abs. Dist. Min.} & \textbf{Sum Abs. Dist. Max.}\\ \hline
        QuKAN (2 layers) &0.7524 &0.5451 &0.0091 &3.3094 \\  \hline 
        QuKAN (1 layer)&0.8332 &0.7157 &0.0142 &3.714 \\  \hline 
    \end{tabular}
    \caption{Average, median, minimum and maximum of the sum of absolute distances for the function regression for the hybrid QuKAN. We chose the training and test set to be of size 250. }
    \label{table1}
\end{table}
\begin{table}[!ht]
    \centering
    \begin{tabular}{|l|l|l|l|l|}
    \hline
        \textbf{Model} & \textbf{Sum Abs. Dist. Avg.} & \textbf{Sum Abs. Dist. Med.} & \textbf{Sum Abs. Dist. Min.} & \textbf{Sum Abs. Dist. Max.}\\ \hline
        QuKAN (2 layers) &0.6833 &0.554 &0.0015 &3.3328 \\  \hline 
        QuKAN (1 layer)&0.7437 &0.5836 &0.0008 &3.7846 \\  \hline 
        FQuKAN &0.995 &0.7542 &0.0079&3.6821 \\  \hline 
        EVQKAN &1.229062 &1.319659 &0.753301 &1.646876 \\  \hline 
    \end{tabular}
    \caption{Comparison of the Average, median, minimum and maximum of the sum of absolute distances between the EVQKAN and (hybrid and fully) QuKAN for a function regression task. The values for the EVQKAN are taken from the Quantum KAN paper by Wakaura et al\cite{wakaura2025enhanced}. Here the train set has 10 samples and the test set has 50 as in Wakaura et al\cite{wakaura2025enhanced}. }
    \label{table4}
\end{table}

\subsubsection*{Pre-training has an effect}
\begin{figure}[!h]
    \centering
    \includegraphics[width=0.9\linewidth]{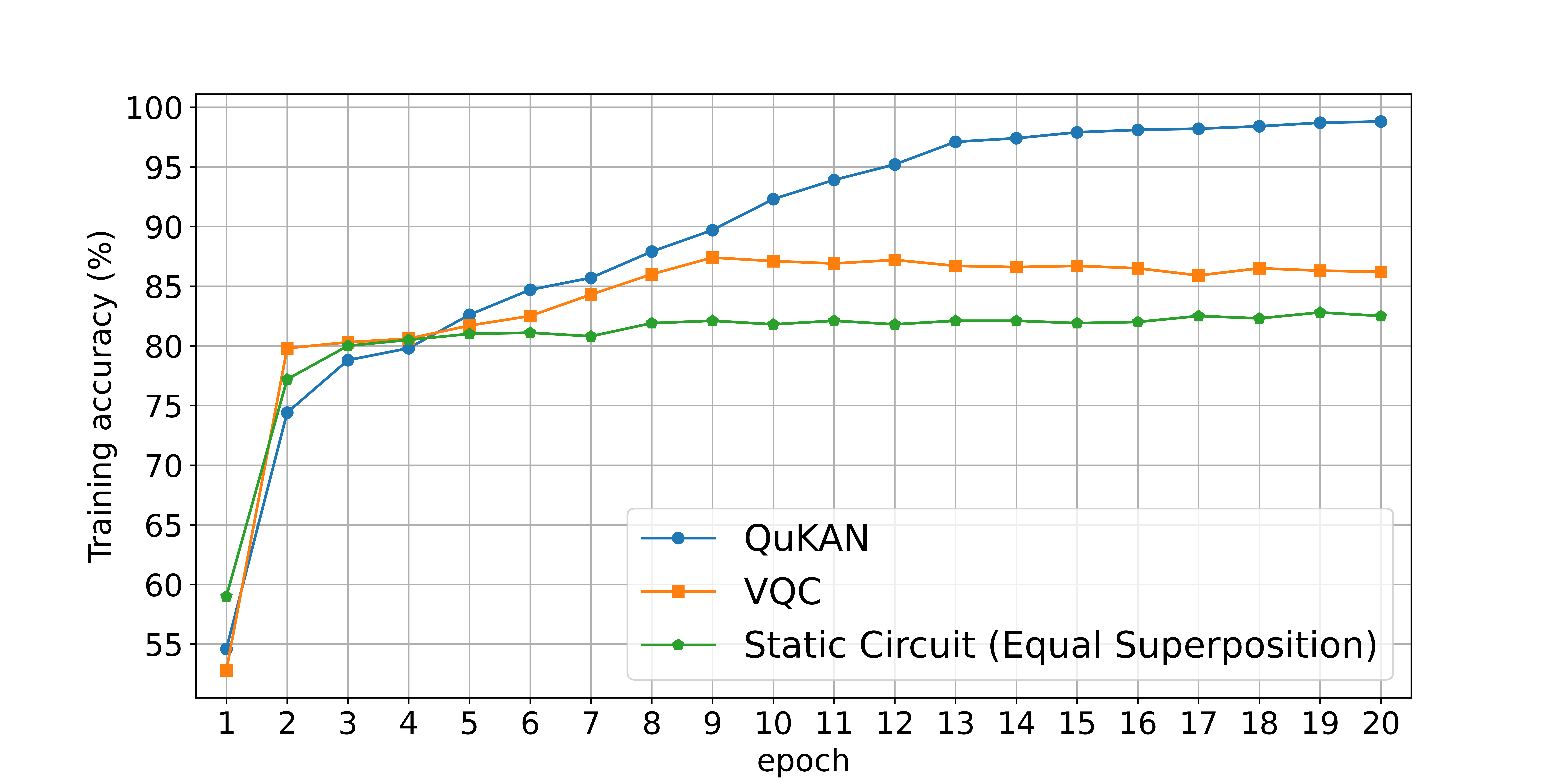}
    \caption{Comparison of the QuKAN training accuracy with a VQC architecture and a random number generator. }
    \label{Fig:Comp}
\end{figure}

To analyse the effect of the pre-training of the QCBM encoded B-Spline basis functions, we perform the training of the binary classification of the moons dataset again. We compare the training behaviour of a VQC with a circuit of equal size as the residual functions to two versions of the QuKAN model: one with pre-trained embedded splines in the quantum residual functions, and one without pre-training initialized in an equal superposition state over the whole computational basis (Hadamard gates). While the output is still given by a projective measurement of the qubit position register, the architecture of the non pre-trained network is analogous to a Variational Quantum Classifier that includes ancillas. We compare the training accuracy over 20 epochs for all models over the moons dataset with noise level $0.1$. The result is presented in Fig.~\ref{Fig:Comp}.
Though both models seems to have an initial success in their learning, the accuracy of the QuKAN without pre-training quickly converges $87.6\%$, indicating that the model does rely on the splines encoded into the QCBM superposition state. We also investigate the effect of removing trainability of the residual function by removing parametrization on the quantum part and replacing by Hadamard gates applied on the complete computational basis. In this case the residual function is equal to a scalable SiLU function and a bias term.

\subsubsection*{Summary of the results}
The presented benchmarks demonstrate that the hybrid Quantum Kolmogorov Arnold Network (QuKAN) is a viable implementation of the classical KAN. On binary classification such as the moons and the Iris datasets, QuKAN achieves high accuracies, outperforming Variational Quantum Classifiers with different embedding strategies. In comparison to classical methods such as MLP and KAN, it remains competitive in terms of interpretability over the number of parameters included for training. Additionally, we compared the classification performance to the QKAN proposed by Ivashkov et al\cite{ivashkov2024qkan}, where we added an autograd ADAM based learning algorithm to their transfer function.

For the function regression task, we could show that the QuKAN trains to fit a linear and a non-linear function. We also showed comparison to the EVKAN\cite{wakaura2025enhanced} model. For the non-linear function we provided data indicating that the fully quantum architecture of the QuKAN residual function in the network is able to train on non linear regression tasks.

Finally, the ablation study on pre-training confirms its crucial role: models trained without the pre-trained QCBM-encoded spline basis quickly plateau during training. This emphasizes that the embedding into quantum residuals prior to training enhances the model's learning capabilities.

\section*{Discussion}
In this work, we have demonstrated that the classical Kolmogorov Arnold Network as proposed by Liu et al \cite{liu2024kan}, can be effectively implemented into a quantum framework using a Quantum Circuit Born Machines approach. We encoded B-Spline basis functions into the amplitude structure of quantum states, allowing them to be interpreted probabilistically via projective measurements of the computational basis states. While we adopted B-Splines to remain consistent with the original KAN formulation, our method is not restricted to a specific choice of basis functions. In principle, any trainable basis set could be used.

By partitioning the qubit register into labelling and position subspaces, we enabled the encoding of multiple basis functions in quantum superposition, with tunable weighting across the function space. These weights are controlled by trainable parametrized unitaries, implemented via strongly entangling layers. This approach allows us to construct both a hybrid quantum KAN and a fully quantum residual analog, while preserving the compositional interpretability that distinguishes KAN from black-box neural networks \cite{xu2024kolmogorov,ranasinghe2024ginn}.
\\ \\
One of the key advantages of our quantum formulation is that it retains interpretability while exploiting quantum superposition to reduce parameter complexity. Specifically, the ability to evaluate multiple basis functions simultaneously through projective measurements results in a compact model. \\ \\
We validated the proposed architecture on toy classification and regression tasks, including moons and Iris datasets as well as multivariate function approximation. We compared the QuKAN to classical as well as quantum methods showing the mean test accuracy over different seeds, thereby showcasing its stability over parameter initialization. Our results suggest that QuKAN provides a promising blueprint for the implementation of KANs into a quantum framework.

\section*{Data availability}
Data and Code will be made available on reasonable request. Correspondence and requests for materials should be addressed to M.K-E.

\newpage


\section*{Acknowledgements}

We gratefully acknowledge financial support from the Quantum Initiative Rhineland-Palatinate QUIP and the Research Initiative Quantum Computing for AI (QC-AI).

\section*{Supplementary Material}
\begin{figure}[!h]
    \centering
    \includegraphics[width=1.0\linewidth]{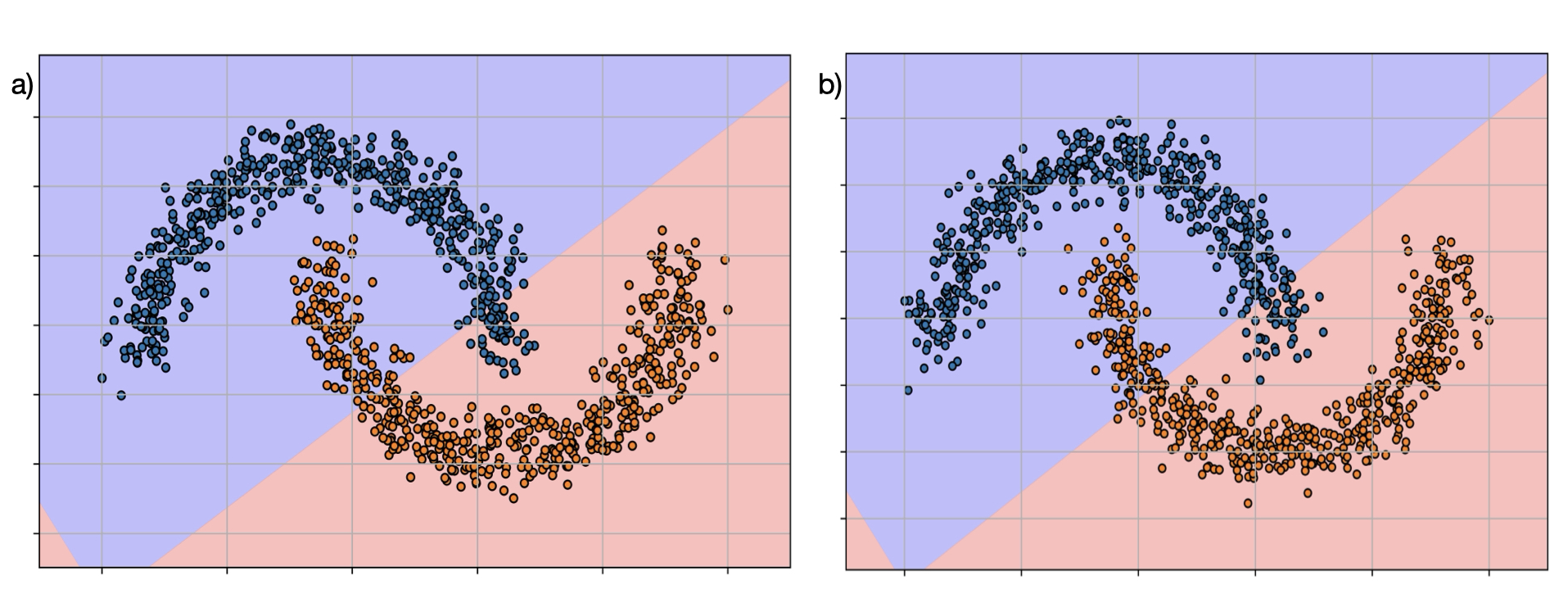}
    \caption{Decision boundaries of the Variational Quantum Classifier with different setups: a) Amplitude Embedding, b) Amplitude Embedding including a total of 4 ancillas. The dataset is the moons dataset with a noise of 0.1. }
    \label{fig:enter-label}
\end{figure}

\begin{table}[h]
    \centering
    \begin{tabular}{lccc}
        \toprule
        \multirow{2}{*}{\textbf{Model}} &
        \multicolumn{3}{c}{\textbf{Noise level}} \\
        \cmidrule(lr){2-4}
        & 0.2 & 0.3 & 0.5 \\
        \midrule
        QuKAN                       & \textbf{93.48\%} & \textbf{89.44\%} & \textbf{83.68\%} \\
        Rigid grid pyKAN                      & 92.40\% & 87.90\% & 79.50\% \\
        VQC (Amplitude Embedding)           & 82.60\% & 82.10\% & 70.10\% \\
        VQC (Amplitude Embedding + Ancillas)       & 83.60\% & 80.90\% & 73.30\% \\
        VQC (ZZ FeatureMap)      & 68.50\% & 69.70\% & 58.00\% \\
        VQC (Angle Embedding)      & 79.70\% & 79.80\% & 53.10\% \\
        \bottomrule
    \end{tabular}
    \caption{Test accuracy of various models trained for 20 epochs and for different noise levels on the moons dataset.}
    \label{tab:best-acc-noise}
\end{table}

\section*{Author contributions statement}
Y.W. performed the implementation and all simulation that include the hybrid and full QuKAN, created the figures and contributed significantly to the writing of the initial manuscript as well as writing the first draft. A.M. implemented the QKAN autograd based learning and performed all VQC related simulations. M.L. contributed in the summary of the classical KAN as well as all simulations regarding pyKAN. V.F.R., N.P., P.L. and M.K-E. provided key insights for the interpretation of the gathered data throughout the study and revised the manuscript and supervised the study. All authors thoroughly reviewed the manuscript and approved it for publication. 

\section*{Additional information}
\noindent \textbf{Declaration of Interest}: The authors declare no competing interests.

\noindent \textbf{AI Tools} like chatGPT4o, Writefull(TeXGPT) have been used only to enhance spelling, grammar as well as shortening long sentences. All improved text has been carefully read and made sure to be free of hallucinations.

\end{document}